\documentclass[10pt,journal]{IEEEtran}
\usepackage{times}
\usepackage{amsmath}
\usepackage{multirow}
\usepackage{graphicx}
\usepackage[usenames,dvipsnames]{color}
\usepackage{subfigure}
\usepackage{parskip} 
\usepackage{algorithm}
\usepackage{algorithmicx}
\usepackage{algpseudocode}
\usepackage{overpic}
\usepackage{verbatim}
\usepackage{multirow}
\usepackage{url}
\usepackage{amssymb}

\definecolor{gold}{rgb}{0.85,.66,0}

\ifCLASSOPTIONcompsoc
  \usepackage[nocompress]{cite}
\else
  \usepackage{cite}
\fi
\ifCLASSINFOpdf
\else
\fi

\begin{document}
%
\title{Deep Online Video Stabilization}
%
%
\author{Miao~Wang,~\IEEEmembership{Member,~IEEE,}
        Guo-Ye~Yang,~\IEEEmembership{Member,~IEEE,}
        Jin-Kun~Lin,~\IEEEmembership{Member,~IEEE,}
                Ariel~Shamir,~\IEEEmembership{Member,~IEEE Computer Society,}
        Shao-Ping~Lu,~\IEEEmembership{Member,~IEEE,}
        and~Shi-Min~Hu,~\IEEEmembership{Member,~IEEE,}
        }

\IEEEtitleabstractindextext{%
\begin{abstract}
Video stabilization technique is essential for most hand-held captured videos due to high-frequency shakes. Several 2D-, 2.5D- and 3D-based stabilization techniques are well studied, but to our knowledge, no solutions based on deep neural networks had been proposed. The reason for this is mostly the shortage of training data, as well as the challenge of modeling the problem using neural networks. In this paper, we solve the video stabilization problem using a convolutional neural network (ConvNet). Instead of dealing with offline holistic camera path smoothing based on feature matching, we focus on low-latency real-time camera path smoothing without explicitly representing the camera path. Our network, called \emph{StabNet}, learns a transformation for each input unsteady frame progressively along the time-line, while creating a more stable latent camera path. To train the network, we create a dataset of synchronized steady/unsteady video pairs via a well designed hand-held hardware. Experimental results shows that the proposed online method (without using future frames) performs comparatively to traditional offline video stabilization methods, while running about 30$\times$ faster. Further, the proposed \emph{StabNet} is able to handle night-time and blurry videos, where existing methods fail in robust feature matching.
\end{abstract}

}

\maketitle

\IEEEdisplaynontitleabstractindextext

%
\IEEEpeerreviewmaketitle

\section{Introduction}

\begin{figure}[t]
\begin{center}
\includegraphics[width=\linewidth]{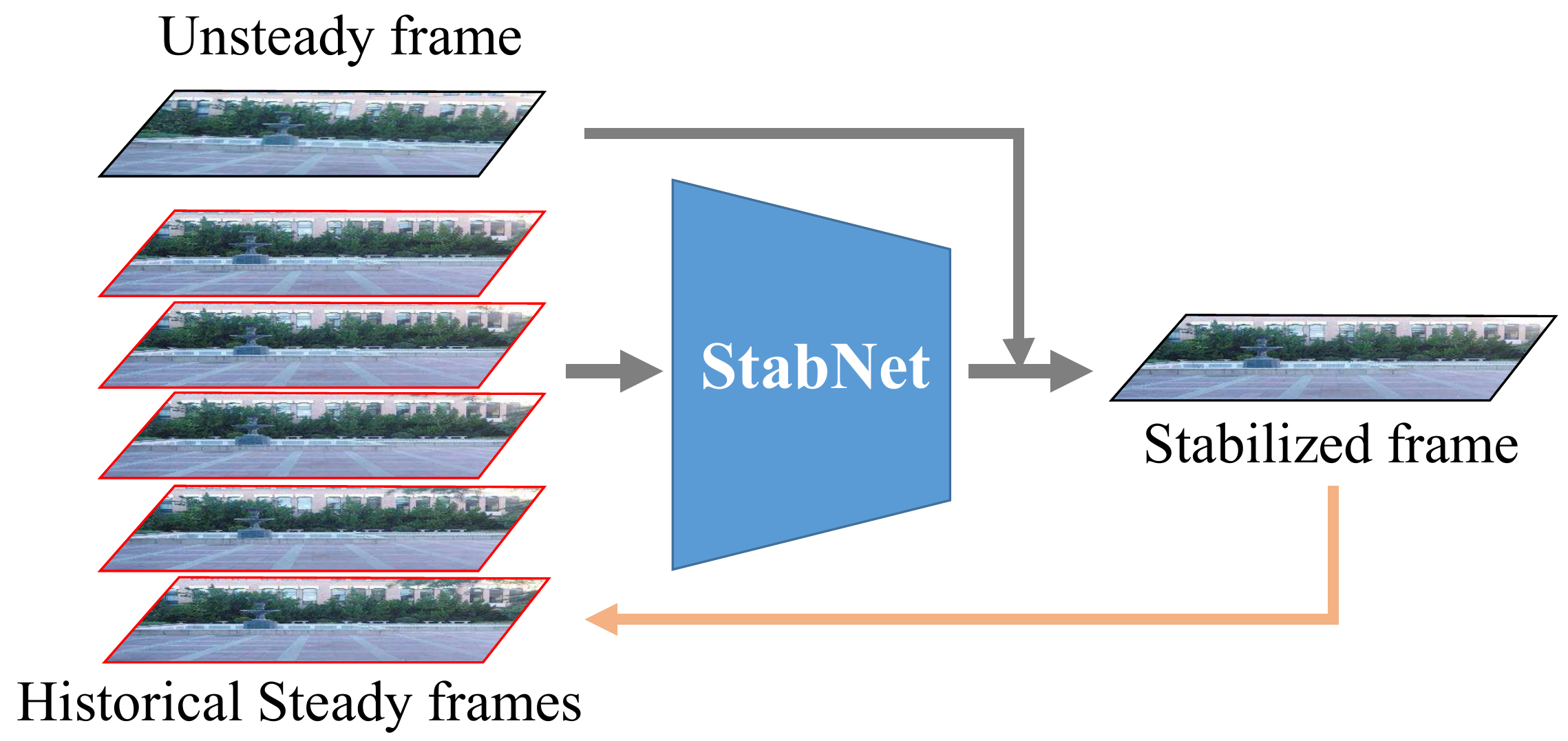}
\end{center}
   \caption{Deep online video stabilization. We propose \emph{StabNet}, a ConvNet model that learns to predict transformation parameters for each incoming unsteady frame, given the history of steady frames. Applying the predicted transformation parameters to the original unsteady frame generates the stabilized output frame. Stabilized frames then act as historical frames for stabilizing future unsteady frames.}
\label{fig:stabnet}
\end{figure}

Video captured by hand-held camera is often not easy to watch due to shaky content. Several digital video stabilization techniques have been proposed in the past decade to improve the visual quality of hand-held videos, by removing high-frequency camera movements \cite{matsushita2006full,liu2009content,grundmann2011auto,liu2011subspace,Bundle}. The majority of the proposed methods deal with this problem using a global view, by estimating and smoothing the camera path using offline computation. The very few online stabilization methods do a `capture$\rightarrow$ compute$\rightarrow$display' operation for each incoming video frame in real time with low latency. Due to the real-time requirement, the camera motion is estimated by an Affine transformation, homography or using meshflow. In this paper, we focus on the online stabilization problem. Different from existing approaches, that must explicitly model the camera path to smooth it, we use a learning-based framework to directly compute a target steady transformation, with guidance from historical stabilized frames (see Figure~\ref{fig:stabnet}).

In recent years, we have witnessed how convolutional neural networks (ConvNets) change vision and graphics fields. In general, methods that are based on ConvNets perform more efficiently. Several traditional video processing topics such as video stylization \cite{huang2017real} and video deblurring \cite{su2016deep} are re-addressed using ConvNets.  To our knowledge, there are no ConvNet-based methods published for digital video stabilization, although it is an important topic in video processing. We observed two main obstacles preventing a ConvNet-based stabilization solution: 1) Lack of training data: pairs of steady and unsteady synchronized videos with an identical capturing route and content are required for training a ConvNet model. While this is not necessary for traditional methods, it is essential for a learning-based stabilization approach. 2) Problem definition: traditional stabilization methods compute and smooth camera path, which cannot be easily adapted to a ConvNet-based solution. A somewhat different problem definition is required. 

Based on these observations, we propose to solve corresponding issues by creating a practical data set for training a neural network, and modifying the formulation of the problem using progressive online stabilization. To collect training data, we captured synchronized hand-held unsteady/steady video pairs using a remodeled hand-held stabilizer with two cameras. With this hardware, only one camera is stabilized by the hand-held stabilizer while the other camera is fixed to the stabilizer grip, moving consistently with the hand motions. In our modified formulation of the stabilization problem, instead of estimating and smoothing a virtual camera path, we learn the transformation parameters for each unsteady frame progressively along the time-line, and generate a steady output video in an online mode. 

We present \emph{StabNet}, a ConvNet model to stabilize frames with light-weighted feed-forward operations through the network. The learning process is driven by the information of historically stabilized frames with the supervised ground-truth steady frame.  Figure \ref{fig:stabnet} shows the overview of  our deep video stabilization.

The proposed deep stabilization method performs comparably well on test videos collected from existing works. The main merit of our algorithm is the ability to run in real-time at 93 FPS with low latency (1 frame), being about 30$\times$ faster than offline methods. Further, our method is superior to existing methods with the ability of handling night-time videos and extreme blurry videos, where existing feature matching based methods may totally fail. To our knowledge, the proposed \emph{StabNet} is a pioneer in using convolutional network for digital video stabilization. We also built the \emph{DeepStab} dataset consisting of pairs of synchronized steady/unsteady videos for training.
We believe that the first stabilization training dataset will benefit the community, and plan to release it for future research.

\section{Related Work}
Our work is closely related to digital video stabilization approaches and deep learning video manipulation.

\paragraph{Digital Video Stabilization} Existing \emph{offline} stabilization techniques estimate the
camera trajectory from 2D, 2.5D or 3D perspective and then synthesize a new smooth camera trajectory to remove the undesirable high-frequency motion. 2D video stabilization methods
estimate (bundled) homography or affine transformations between consecutive frames and smooth these transformations temporally.
In early work, low-pass filters were applied to individual
model parameters \cite{matsushita2006full,CHANG2006659} . Grundman et al. applied L1-norm
optimization to synthesize a path consisting of simple cinematography motions \cite{grundmann2011auto}. Liu et al. \cite{Bundle} modeled the camera motion on multiple local camera paths. Zhang et al. \cite{Zhang:2017:GVS:3083730.3083944} proposed to optimize geodesics on the Lie group embedded in transformation space. 3D-based
stabilization methods reconstruct a 3D scene \cite{snavely2006photo} and estimate
the 3D camera trajectory. Liu et al. \cite{liu2009content} proposed the first 3D
stabilization method using content-preserve warping. The later
subspace video stabilization method \cite{liu2011subspace} smooths long tracked
features using subspace constraints. Goldstein and Fattal
\cite{Goldstein:2012:VSU:2231816.2231824} proposed to enhance the length of feature tracks with epipolar transfer. The above two 2.5D based methods can deal with cases of
reconstruction failure, and produces results that are visually
as good as a 3D-based method. With user interactions, Bai et al. \cite{Bai:2014:UVS:2855536.2855546} proposed a user-guided stabilization approach to select good feature tracks and warping results. Rolling shutter problem in high-speed video is addressed in \cite{37744}. Recently, a 2D-3D hybrid
stabilization approach was proposed to stabilize 360 video
\cite{360stabilization}. In general, 2D stabilization methods perform efficiently
and robustly, while 3D-based methods can generate visually
better results.

Real-time \emph{online} stabilization is specifically desired for live stream applications. Liu et al. \cite{Meshflow} proposed an online stabilization method which only use historical camera path to compute warping functions for incoming frames. Inspired by their idea, we present a deep online stabilization approach which performs stabilization given afew historical stabilized frames. The novelty of our approach is that we avoid explicitly estimating and smoothing camera path, instead, we use a ConvNet model to directly predict warping functions.

\paragraph{ConvNets for Video Applications} In recent years, ConvNets have made huge improvements in computer vision tasks such as image recognition \cite{Krizhevsky:2012:ICD:2999134.2999257,DBLP:journals/corr/SimonyanZ14a,he2016deep}, segmentation \cite{long2015fully,zhao2016pyramid,he2017mask} and generation \cite{gatys2015neural,isola2016image,DBLP:journals/corr/ZhuPIE17}. When feeding multiple successive frames from videos, ConvNets can predict optical flow \cite{dosovitskiy2015flownet,IMKDB17} or semantics \cite{simonyan2014two,diba2016deep}. There are several works which use ConvNets to directly produce video contents, such as scene dynamic generation \cite{Vondrick2016GeneratingVW,visualdynamics16}, frame interpolation \cite{liu2017video,DBLP:journals/corr/abs-1708-01692} and deblurring \cite{su2016deep,kim2017online}. Because predicting a long video sequence is still a challenging problem, all of the above works used only two or very few successive frames as training samples. The proposed StabNet also considers a temporal neighborhood at each time. The stabilization problem cannot be solved using a generation-based model because of the severe vibration of the input video content. To generate visually pleasing result, our \emph{StabNet} learns the warping parameters instead of generating pixel values.
\section{Training Dataset}
\label{sec:trainingset}
\begin{figure*}[t]
\begin{center}

\includegraphics[width=\linewidth]{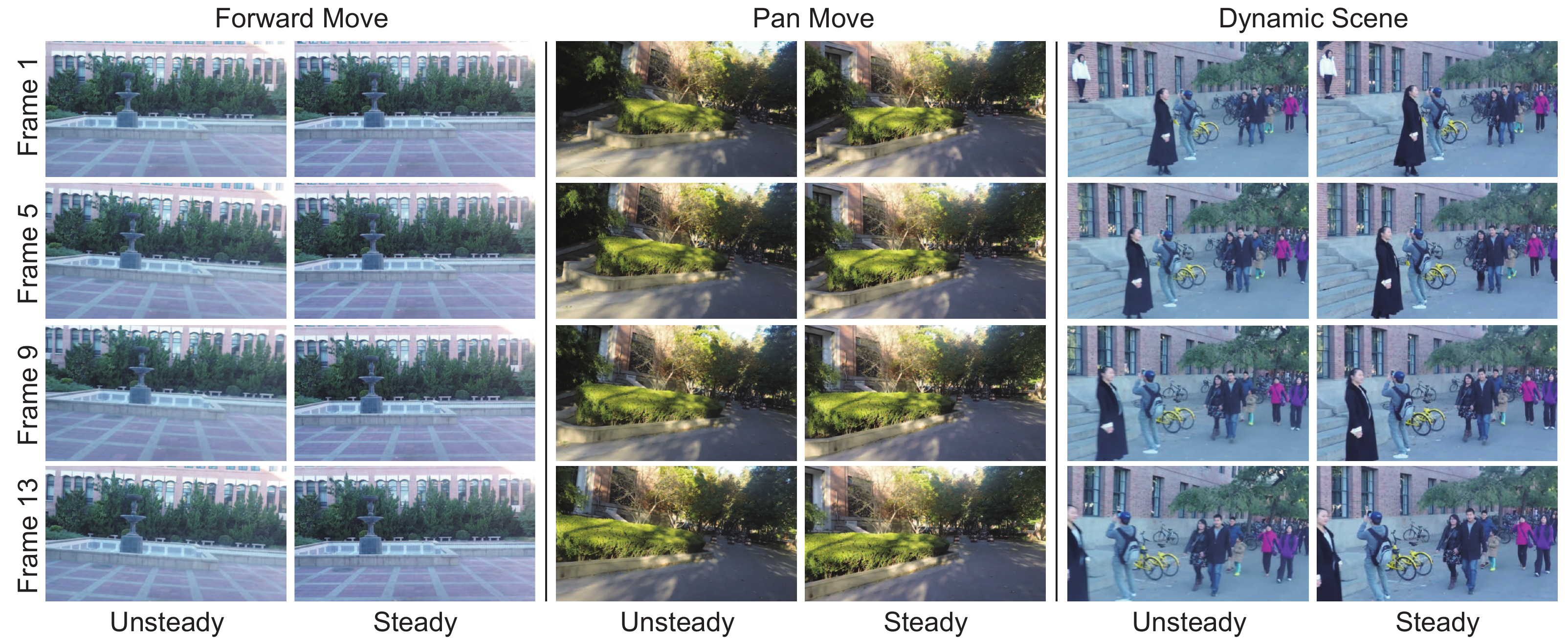}
\end{center}
   \caption{Exemplar frames of DeepStab dataset. The dataset includes pairs of synchronously captured videos. Each pair consists of an unsteady video and a stabilized video, with the same content. Camera motions include forward movement, pan movement, spin movement and complex movements including combinations of the above, at various speed.}
\label{fig:dataset}
\end{figure*}
\begin{figure}[t]
\begin{center}
\includegraphics[width=0.9\linewidth]{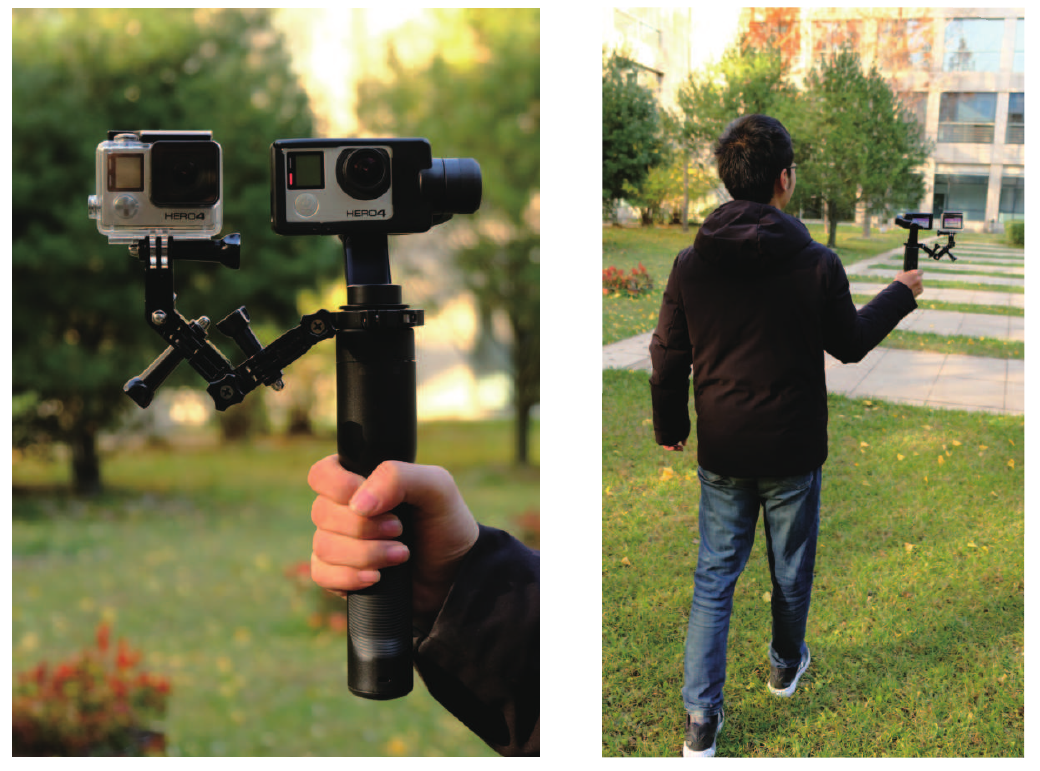}
\end{center}
\caption{Hardware and training data capturing process.}
\label{fig:hardware}
\end{figure}

Generating training data is one of the key challenges
for digital video stabilization, where ground truth data cannot be easily collected/labeled.
To train StabNet, two synchronized video sequences of the same scene are required:
one sequence captures a steady camera movement, while the other is unstable. One possible way to generate such data is to
render a virtual scene with two camera path configurations: smooth and turbulent. 
However, ConvNet models trained using rendered virtual scene may not
generalize well due to the domain gap between training synthetic video and testing real videos captured by hand-held camera. 
To generate authentic data, we designed a hardware with two portable GoPro Hero 4 Black cameras and a hand-held stabilizer\footnote {\url{https://www.youtube.com/watch?v=8vu7IDuDD64}}, where the cameras lay horizontally next to each other with small disparity (Figure \ref{fig:hardware}). When capturing videos, the two cameras shoot synchronously, with only one camera stabilized, while the other moves consistently with the hand/body motion of the holder. We turned off the auto-focus and auto-exposure functions
of the cameras and used the synchronous remote control for synchronization.

Training videos are obtained by holding the designed hardware while taking shots in a first-person point of view. We present the \emph{DeepStab} dataset, containing pairs of synchronized videos of outdoor scenes with diverse camera movements. The dataset includes indoor scenes with large parallax and common outdoor scenes with buildings, vegetation, crowd, etc. Camera motions include forward movement, pan movement, spin movement and complex movements including combinations of the above, at various speed. Videos are processed to remove the fish-eye distortion and trim out clips with large lighting difference. 

In total, we collected 60 pairs of synchronized videos, each is 20-30 seconds long at 30 FPS on average. The videos are split into 44 training pairs, 8 validation pairs and 8 testing pairs. Figure \ref{fig:dataset} shows representative sampled frames from the dataset. The recorded video pairs are augmented to provide more training samples by horizontally flipping the frames, reversing the video sequences and combining both flipping and reversing.


\section{The StabNet}
\label{sec:stabnet}
\begin{figure*}[t]
\begin{center}
\includegraphics[width=\linewidth]{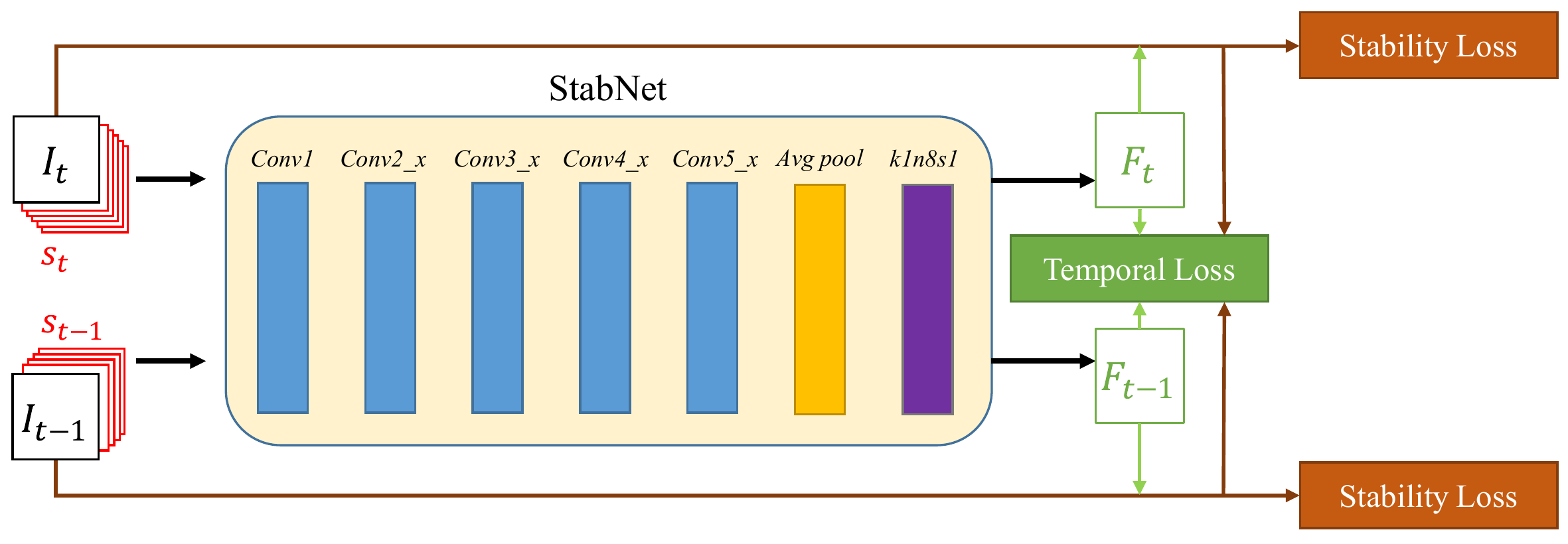}
\end{center}
\caption{Network Architecture. \emph{StabNet} is a two-branch Siamese network with shared parameters in each branch. It consists of an \emph{encoder} and a \emph{Homography Regressor}. Homography Regressor is a Conv layer output 8 channels with 1$\times$1 kernel size and 1 pixel stride (k1n8s1). During training, samples of two successive frames $\langle I_t, s_t\rangle$ and $\langle I_{t-1}, s_{t-1}\rangle$ are fed to the network, and the transformation parameters $F_t$ and $F_{t-1}$ predicted. The network is trained with both a \emph{stability loss} and a \emph{temporal loss}.}
\label{fig:pipeline}
\end{figure*}
\textbf{Overview:} 
As our focus is online stabilization, we cannot use future frames when processing a given frame. We convert the online stabilization problem to a supervised learning problem of conditional transformation without explicitly computing a camera path. The inputs to \emph{StabNet} are an incoming unsteady frame $I_t$ and conditionally 5 historical steady frames sampled from approximately one previous second $S_{t}=\langle \dot{I}_{t-30}, \dot{I}_{t-24}, \dot{I}_{t-18}, \dot{I}_{t-12}, \dot{I}_{t-6} \rangle$ for time-stamp $t$. The output is a transformation $F_t$ for frame $I_t$. The steady frame is created by applying $\widehat{I}_t = F_t\ast I_t$, where $\ast$ is the warping operator. The learning process is supervised by ground-truth steady frames $I'_t$. When training \emph{StabNet}, the conditional inputs $S_{t}$ used are the ground-truth steady frames $\langle {I}'_{t-30}, {I}'_{t-24}, {I}'_{t-18}, {I}'_{t-12}, {I}'_{t-6} \rangle$, while for testing, $S_{t}$ are the historical stabilized frames $\langle \widehat{I}_{t-30}, \widehat{I}_{t-24}, \widehat{I}_{t-18},  \widehat{I}_{t-12},  \widehat{I}_{t-6}\rangle$.


\subsection{Network Architecture} 
 Our \emph{StabNet} is a Siamese network \cite{zagoruyko2015learning} that has two branches sharing the network parameters. 
We use a Siamese architecture to preserve temporal consistency of successive transformed frames $\widehat{I}_{t-1}=F_{t-1}\ast I_{t-1}$ and $\widehat{I}_{t}=F_t\ast I_t$. Each branch of \emph{StabNet} is a two-stage network consisting of an \emph{encoder}, that extracts high-level features from the inputs and a \emph{regressor}, that predicts the stabilization transformation parameters from the extracted feature map. Figure \ref{fig:pipeline} shows the architecture of \emph{StabNet}. The inputs are 6 concatenated grayscale frames, each with dimension $W\times H\times 1$, consisting of 5 conditional steady frames $S_{t}$ and one unsteady frame $I_t$. Frames are sent to an \emph{encoder} to extract features. This encoder adapts ResNet-50 \cite{he2016deep} as the backbone feature extractor, using the \emph{conv\underline{ }1} as the input channel, modified to meet our inputs, and removing all layers after \emph{average pooling}. The extracted feature map from the \emph{encoder} is of dimension $1\times1\times2048$. Next, we use a conv layer to regress a Homography transformation with $1\times1$ kernel size. 
A Homography transformation has eight parameters to regress:
\begin{equation}
F = 
\left(
 \begin{matrix}
   h_1 & h_2 & h_3 \\
   h_4 & h_5 & h_6 \\
   h_7 & h_8 & 1
  \end{matrix}
  \right),
\end{equation}
these parameters could be represented using a $1\times8$ vector. As a result, a $1\times1\times8$ vector is produced as the output of \emph{StabNet}.

\subsection{Stabilization Loss Function}
\label{SLF}
\emph{StabNet} training process is driven by a two-terms loss function including a stability term and a temporal smoothness term, which is based on neighboring unsteady frames $I_t$ and $I_{t-1}$ The loss function is defined as:
\begin{equation}
L =\sum_{i\in\{t,t-1\}}  L_{stab}(F_i, I_i) +\lambda L_{temp}(F_t, F_{t-1}, I_t, I_{t-1}),
\end{equation}
where $L_{stab}$ is the \emph{stability loss}, and $L_{temp}$ is the \emph{temporal loss}. $\lambda=30.0$ is a weighing parameter balancing the two losses. 

\subsubsection{Stability Loss}
The \emph{stability loss} drives the warped unsteady frames to the ground-truth steady frames using cues of pixel alignment and feature point alignment. 
It is defined as:
\begin{equation}
L_{stab}(F_t, I_t)=L_{pixel}(F_t, I_t) + \alpha L_{feature}(F_t, I_t),
\end{equation}
where $L_{pixel}$ is the pixel alignment term, $L_{feature}$ is the feature alignment term, and $\alpha$ is a weighing parameter set to $0.33$.

The pixel alignment term $L_{pixel}$ measures how the transformed frame $\widehat{I}_t = F_t\ast I_t$ aligns with the ground-truth steady frame ${I}'_t$, using mean squared error (MSE):
\begin{equation}
L_{pixel}(F_t, I_t)=\frac{1}{D}||{I}'_t - F_t\ast I_t||^2_2,
\end{equation}
where $D$ is the spatial dimension of frame.
The transformation $F_t\ast I_t$ operates in the image domain. To make the warping function differentiable, we used spatial transformer layer \cite{jaderberg2015spatial}.
$L_{pixel}$ loss will be small if the transformed frame $\widehat{I}_t$ aligns well with the ground-truth frame ${I}'_t$. However, during training  $\widehat{I}_t$ can converge slowly to ${I}'_t$. During early training stages, frames are not aligned well and the loss term is less correlated. For faster convergence during training, we further introduce a feature alignment loss.

The feature alignment term $L_{feature}$ is computed as the average alignment error of matched \emph{feature points} after transforming the unsteady frame $I_t$ using the predicted transformation $F_t$:
\begin{equation}
L_{feature}(F_t, I_t)=\frac{1}{m}\sum_{i=1}^m||p'^i_t - F_t\ast p^i_t||^2_2.
\end{equation}
where $P_t=\{\langle p^i_t, p'^i_t \rangle \; |\; i\in\{1, \cdots, m\}\}$ are the $m$ pairs of matched feature points between each steady/unsteady frame pair, and $p^i_t$ and $p'^i_t$ are the $i$-th matched feature points from unsteady frame $I_t$ and ground-truth steady frame $I'_t$ respectively.

To compute the feature loss, all pairs $P_t$ are computed in a pre-processing stage between steady and unsteady frame pairs. We extract SURF features \cite{Bay2006} from both $I_t$ and  $I'_t$, then calculate the matching between them by dividing the frames into $2\times2$ sub-images, and using a RANSAC algorithm \cite{Fischler:1981:RSC:358669.358692} to fit a Homography in each corresponding sub-image. We match features in $2\times2$ sub-images instead of $4\times4$ as in \cite{Bundle}, because of the large camera pose and content variation between the steady and unsteady cameras.
Note that the feature extraction and feature matching processes are \emph{only} performed for training the network and not needed during online stabilization.

\subsubsection{Temporal Loss}
\begin{figure}[!t]
\begin{center}
\includegraphics[width=\linewidth]{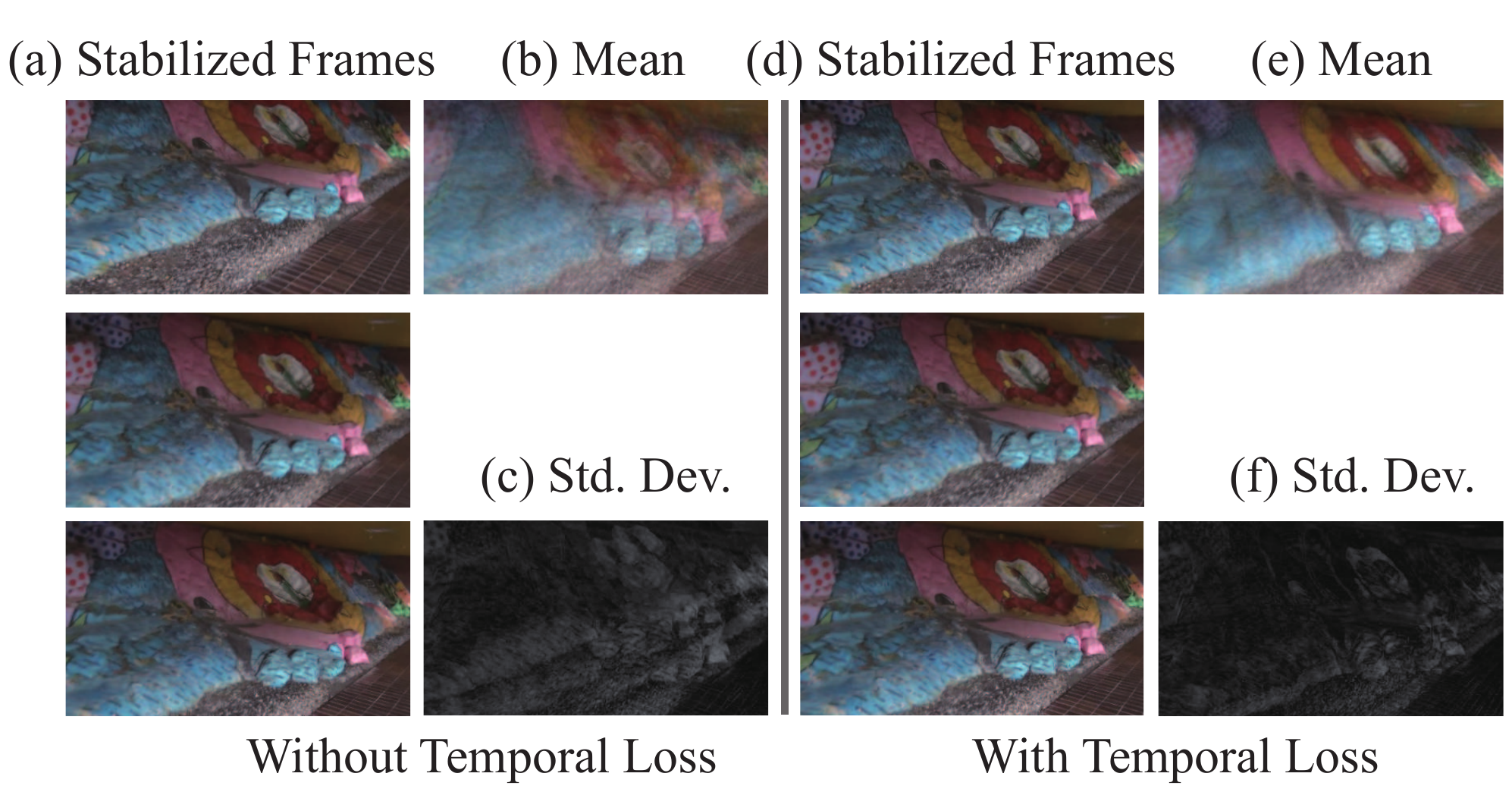}
\end{center}
   \caption{Stabilization result without and with temporal loss. Results with temporal loss leads to sharper, less
“ghosted” mean and lower standard deviations. (a), (b) and (c) are successive stabilized frames, corresponding mean and standard deviation value without temporal loss; (d), (e) and (f) are successive stabilized frames, corresponding mean and standard deviation value with temporal loss.}
\label{fig:temporal}
\end{figure}

Simply applying the transformations separately to every video frame can create wobble artifacts in the video. Therefore, we incorporate a
temporal loss term to enforce temporal coherency between adjacent frames using the Siamese network architecture. Figure \ref{fig:temporal} shows the comparison of stabilization result with
and without temporal loss. Each time two successive samples $\langle I_t, s_{t} \rangle$ and $\langle I_{t-1}, s_{t-1} \rangle$ are fed into \emph{StabNet}, two successive transformations $F_t$ and $F_{t-1}$ are predicted. The temporal loss is defined as the mean square error between the successive output frames:
\begin{equation}
L_{temp}(F_t, F_{t-1}, I_t, I_{t-1})=\frac{1}{D}||F_t\ast I_t - w(F_{t-1}\ast I_{t-1})||_2^2,
\end{equation}
where $D$ is the spatial dimension of frame, $w(\cdot)$ is a function
that warps the steady frame at $t-1$ to the steady frame $t$ according
to pre-computed optical flow. In our experiments we use TV-L1 algorithm \cite{perez2013tv} to compute the optical flow, but alternative methods for optical flow calculation can also be used. 
\begin{figure*}[!t]
\begin{center}
\includegraphics[width=\linewidth]{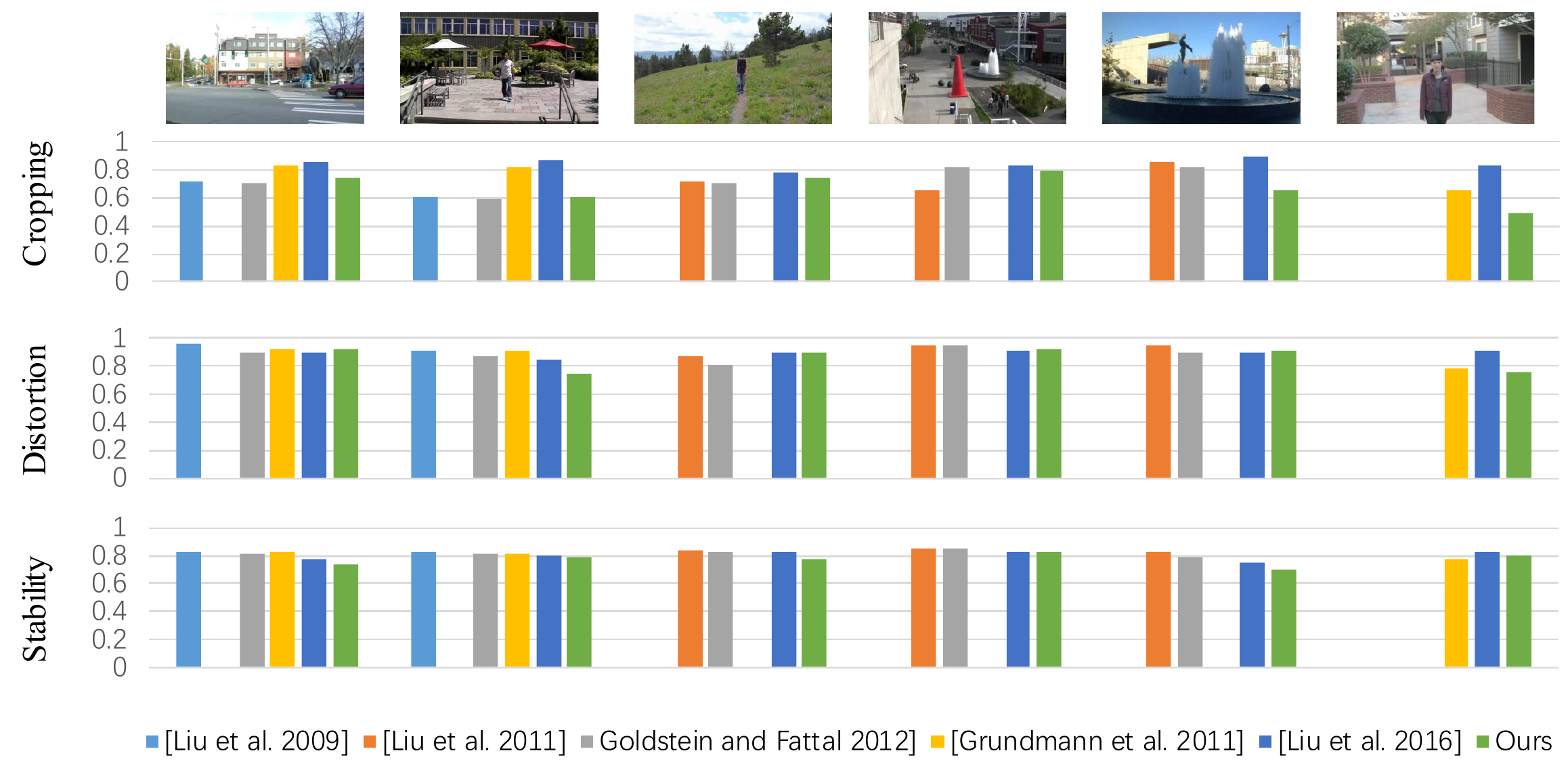}
\end{center}
\caption{Comparison with 6 publicly avaliable videos in terms of three metrics.}
\label{fig:statistic}
\end{figure*}

\subsection{Implementation Details}

To train \emph{StabNet}, we resize the videos to a spatial dimension of $W=512$ and $H=288$ for efficiency. Pre-trained ResNet-50 model on ImageNet \cite{imagenet_cvpr09} without the \emph{Conv\underline{ }1} layer is loaded, and is fine-tuned during the training process. We use mini-batch size of $8$ and ADAM \cite{kingma2014adam} for optimization with $\beta_1=0.9$, $\beta_2 = 0.999$. Initial learning rate is set to $0.001$, and multiplied by $0.1$ every $30,000$ iterations. The training process is terminated when reaching $90,000$ iterations. The whole training process takes about $20$ hours on an NVIDIA GTX 1080 Ti graphics card.

In the training process, we feed \emph{StabNet} with two successive samples in the two branches (with shared network parameters) to learn temporal coherency. However, during testing, the network is used to stabilize a single frame at a time; temporal consistency is automatically preserved. Further, the stabilization is self-driven for a test video as follows: we start by duplicating the first frame $r$ times and regard the duplicated frames as $S_1$. After stabilizing, frame $I_t$, historical stabilized frames $\langle \widehat{I}_{t-29},\widehat{I}_{t-23},\widehat{I}_{t-17},\widehat{I}_{t-11},\widehat{I}_{t-5}\rangle$ are regarded as  $S_{t+1}$ for stabilizing the next frame $I_{t+1}$. This process is repeated through the time-line. 

The stabilization results inevitably have meaningless frame borders introduced by the warping function. As \emph{StabNet} uses stabilized frames as the inputs for future frames, we need to make \emph{StabNet} robust to such borders. During training, we add some black borders produced by Homography perturbation around the Identity transformation to the ground-truth historical frames. The Homography perturbances are randomly sampled between $H_{min}=\left[ \begin{array}{ccc}
0.9 & -0.1 & -0.5\\
-0.1 & 0.9 & -0.5\\
-0.1 & -0.1 & 1
\end{array} 
\right ]$ and $H_{max}=\left[ \begin{array}{ccc}
1.1 & 0.1 & 0.5\\
0.1 & 1.1 & 0.5\\
0.1 & 0.1 & 1
\end{array} 
\right ]$, where the image axis is normalized to $[-1,1]$. For testing, we crop and trim the borders in post-processing. We plan to release source code and pre-trained \emph{StabNet} model.

\begin{figure*}[!t]
\begin{center}
\includegraphics[width=\linewidth]{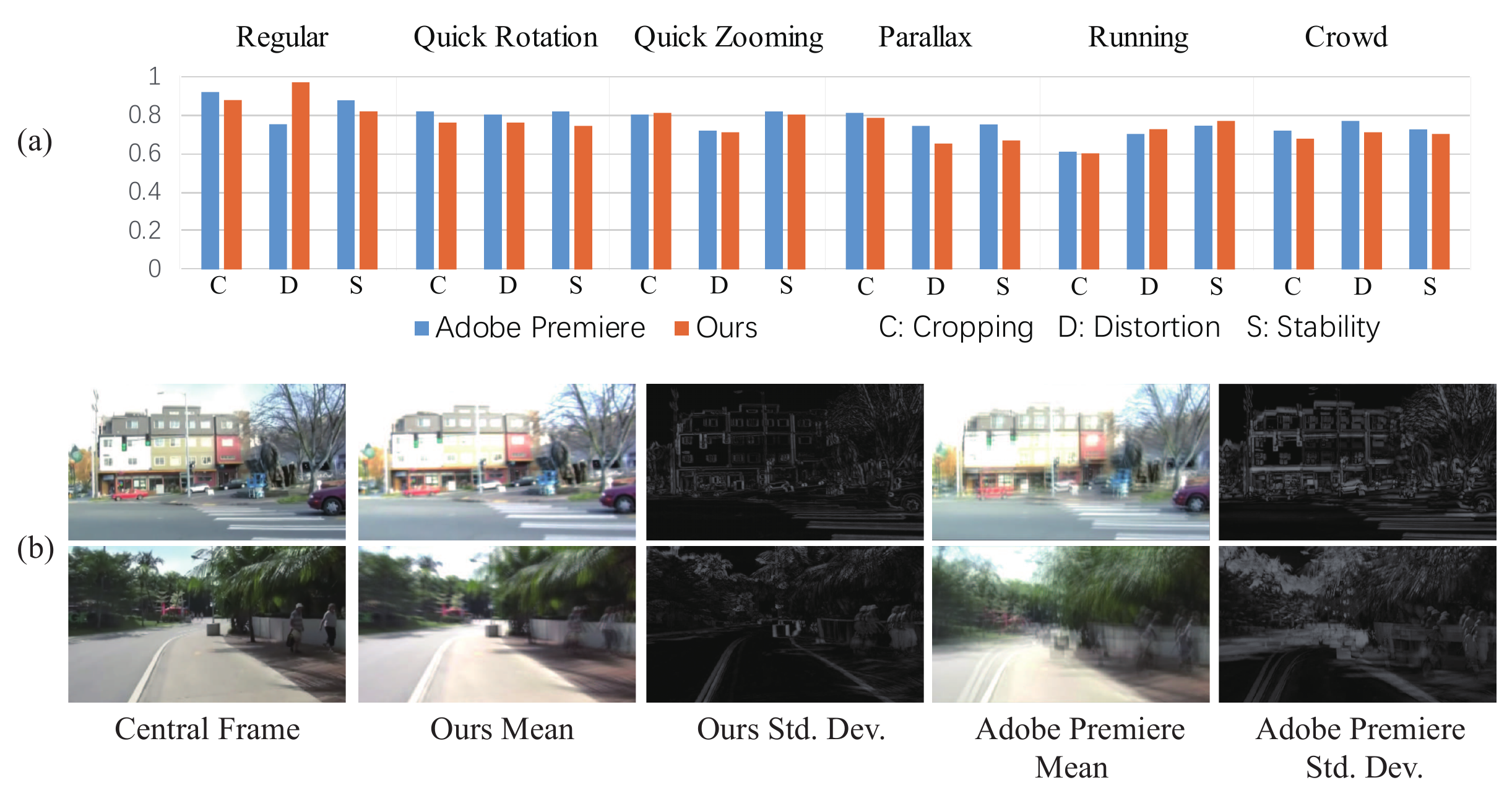}
\end{center}
\caption{Comparison with commercial software Adobe Premiere CS6. (a) Quantitative evaluation; (b) Visual comparison of 5 consecutive stabilized frames with its central frame, average frame and standard deviation.}
\label{fig:evaluation}
\end{figure*}

\section{Experimental Results}
\label{sec:results}

We tested our method on various video sources. Testing videos are either from our \emph{DeepStab} testing set or from \cite{Bundle}. On average, testing runs at $93.3$ FPS, which meets the requirement of real-time online stabilization with $1$ frame latency.

\subsection{Quantitative Evaluation}
Online stabilization problem is inherently harder than offline stabilization, because only historical frames are available in online stabilization, without global sense of the camera path. Hence, quantitative performance statistics for online stabilization methods would be inferior to offline ones.

We compare our method with existing approaches using quantitative evaluation, computed following \cite{Bundle,Meshflow}. The three objective metrics are \emph{cropping ratio}, \emph{distortion} and \emph{stability}.

\paragraph{Cropping ratio} This metric measures the area of the remaining content after stabilization. Larger cropping ratio with less cropping is favored. Per frame cropping ratio is computed as the scale component of the global Homography $H_t$ estimated from input frame $I_t$ to output frame $\widehat{I}_t$. Ratio values of video frames are averaged to generate the cropping ratio value of the whole video.

\paragraph{Distortion value} Distortion value evaluates the distortion degree introduced by stabilization. Per frame distortion value is computed by the ratio of the two largest eigenvalues of the affine part of the Homography $H_t$. The minimum value which represents the worst distortion is chosen as the distortion value for the whole video. 

\paragraph{Stability score} Stability score measures how stable a video is. As there is no benchmark evaluating stabilization videos, following \cite{Bundle}, we use frequency-domain analysis of camera path to estimate the stability score. The camera path is computed as accumulated Homography transformations between successive frames: $\mathcal{P}^t=H_0 H_1\cdots H_{t-1}$. We extract the rotation and translation component from $\mathcal{P}^t$ as a 1D temporal signals, and take each of their lowest frequencies from 2nd to 6th components over full frequencies as translation and rotation stability score. Finally, we take the minimum value of the translation stability and rotation stability as the overall stability score.

We compare 6 publicly available videos against \cite{liu2009content,liu2011subspace,Goldstein:2012:VSU:2231816.2231824,grundmann2011auto,Meshflow} in terms of the objective metrics, based on results provided by corresponding authors. Comparing against offline stabilizations  is slightly unfair for our method because future-frames information is not available for our online stabilization method in real-time. As a result, the stability score of our method is slightly lower. Nevertheless, our method performs in real time while being visually comparable to all existing methods. Comparison details are shown in Figure \ref{fig:statistic}, for videos that
we do not find the result, we leave it blank.

We further compare our method with commercial offline stabilization 
software \emph{Adobe Premiere CS6} on a publicly available video data set \cite{Bundle} and our \emph{DeepStab} data set. As far as we know, Adobe Premiere stabilizer is developed according to the methods in \cite{liu2011subspace}. We choose
the default parameters for Adobe Premiere (smoothness: 50$\%$, ‘Smooth Motion’ and
‘Subspace Warp’) to produce results. The testing sets group videos into several categories according to scene type and camera motion, including \emph{Regular}, \emph{Quick Rotation}, \emph{Quick Zooming}, \emph{Large Parallax}, \emph{Running} and \emph{Crowd}. Evaluation is reported in Figure \ref{fig:evaluation}. It can be seen from Figure \ref{fig:evaluation} (a) that our method generally performs as well as Adobe Premmiere CS6, and from Figure \ref{fig:evaluation} (b) our results would be better than Adobe Premmiere CS6 in \emph{Regular} and \emph{Running} categories.

\paragraph{Discussion} As mentioned, the quantitative evaluation score of our online stabilization method is inevitable lower than offline methods. However the average running time performance of our method is superior to all existing methods. Further, our method is only based on historical frames, thus can be used for online streaming. Running time performance is given in Table \ref{tab:runingtime}. 
\begin{table}
\caption{Running time comparison. FPS statistics are given in the second column. Third column shows whether future frames are required for stabilization.}
\begin{center}
\begin{tabular}{|l|c|c|}
\hline
Method & FPS & Future Frames\\
\hline\hline
Bundled Cameras & 3.5 & Yes\\
Adobe Premiere & 4.2  & Yes\\
MeshFlow & 22.0 & No\\
Ours & 93.3 & No\\
\hline
\end{tabular}
\end{center}

\label{tab:runingtime}
\end{table}
\begin{figure}[t]
\begin{center}
\includegraphics[width=\linewidth]{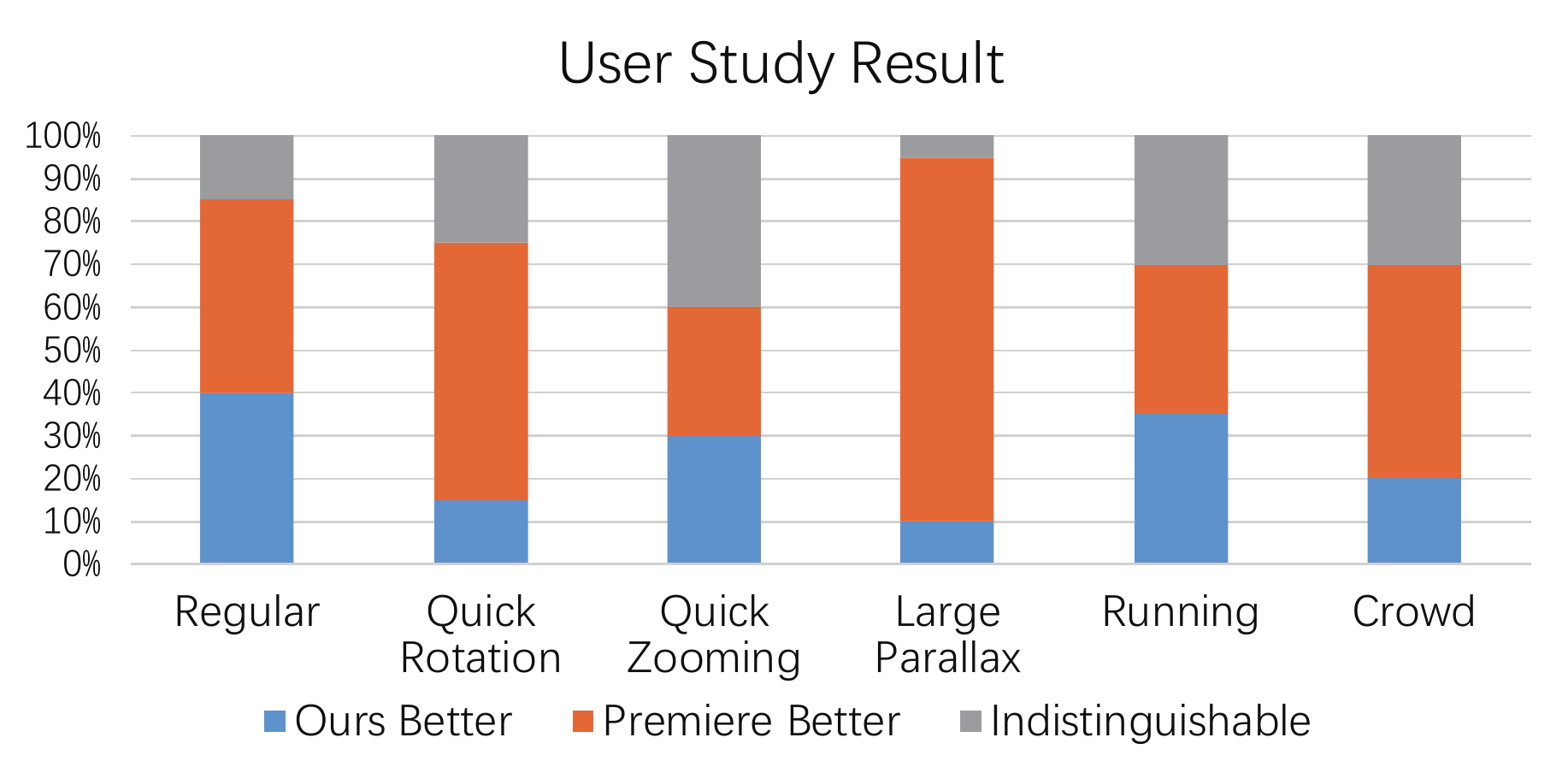}
\end{center}
\caption{User study result by comparing our method with Adobe Premiere  Stabilizer.}
\label{fig:userstudy}
\end{figure}
\subsection{User Study}
To visually compare our method with Adobe Premiere stabilizer, we further conduct a user study with 20 participants aged from 18 to 32.
We provide 18 videos for testing, 3 from each aforementioned category. In each testing case, we simultaneously show
the original input video, our result, and the result from Adobe Premiere stabilizer to the subjects. The two stabilization results are displayed horizontally in random order. Every participant is asked
to pick the more stable result from the results of our method and Adobe Premiere
stabilizer, or mark them `indistinguishable', while disregarding differences in aspect ratio, or sharpness.

The user study results are shown in Figure \ref{fig:userstudy}. For each category, we
show the average percentage of user preference. It can be concluded that for videos from \emph{Regular}, \emph{Quick Zooming}, \emph{Running} categories, our results are comparable with those from offline approach. For other categories that were harder to process without future-frames information, our result is slightly worse. The user study result coincide with our aforementioned discussion.

\subsection{Handling Low Quality Videos}
\emph{StabNet} is robust to low quality videos caused by noise, motion blur or low resolution. When dealing with such kind of videos, traditional methods could fail because few or even no features are available for computing camera path. We show low quality video stabilization results on night-time video and blurry video cases in supplemental video.

\subsection{Limitations}
\emph{StabNet} has limitations. First, in our current implementation, a global Homography transformation is predicted for stabilizing each unsteady frame. More complex network architectures and more complex transformations such as \emph{bundled transformations} can be further explored. Second, in scenes with drastic motion or with extreme near-range foreground objects, our method may fail. We note that these scenarios are also a challenge for previous methods \cite{liu2011subspace,grundmann2011auto,Bundle,Meshflow}.

\section{Conclusion}

We have presented \emph{StabNet}, a convolutional network for digital online video stabilization. Unlike traditional methods which calculate estimated camera paths, \emph{StabNet} learns a warping transformation for each unsteady frame, only using historical stabilized frames as condition. It runs in real time by fast feed-forward operations. We also present \emph{DeepStab} - a dataset consisting of pairs of synchronized steady/unsteady videos for training. This set was created using a practical method to generate training videos with synchronized steady/unsteady frames, which could benefit future deep stabilization methods. To our knowledge, \emph{StabNet} is the first ConvNet for video stabilization. 
We have demonstrated the power of \emph{StabNet} for handling typical types of hand-held videos. We believe ConvNet-based methods are promising for digital video stabilization.

\ifCLASSOPTIONcaptionsoff
  \newpage
\fi



\bibliographystyle{IEEEtran}
\bibliography{egbib.bib}
\end{document}